\documentclass[twocolumn]{aastex631}

\usepackage{ifthen}
\usepackage{xspace}
\usepackage{upgreek}
\usepackage{nicefrac}
\usepackage{bm}
\usepackage{ulem}
\usepackage{xspace}
\usepackage[italicdiff]{physics}

\definecolor{dg}{rgb}{0.0, 0.6, 0.1}
\definecolor{ed}{rgb}{1.0, 0.6, 0.1}

\newcommand{\Andrew}[1]{\textcolor{dg}{#1}}

\usepackage{pgffor}
\makeatletter
\def\dif{\@ifnextchar[{\@with}{\@without}}
\def\@with[#1]#2{
  \ensuremath{\frac{\foreach \x in {#2}{\mathrm{d}\x\,}}{\foreach \x in {#1}{\mathrm{d}\x\,}}}
}
\def\@without#1{
  \ensuremath{%
    \ifx\hfuzz#1\hfuzz
    \mathrm{d}
    \else
    \foreach \x in {#1}{\mathrm{d}\x\,}
    \fi
    }
}
\makeatother

\newcommand{\unit}[1]{\ensuremath{\,\mathrm{#1}}}
\newcommand{\erg}{\ensuremath{\mathrm{erg}}}
\newcommand{\ergs}{\ensuremath{\mathrm{erg\,s^{-1}}}}
\newcommand{\ergscm}{\ensuremath{\mathrm{erg\,s^{-1}\,cm^{-2}}}}

\newcommand{\be}{\begin{equation}}
\newcommand{\ee}{\end{equation}}
\newcommand{\ba}{\begin{eqnarray}}
\newcommand{\ea}{\end{eqnarray}}

\newcommand{\ve}{\ensuremath{\varepsilon}}

\newcommand{\G}{\ensuremath{\Gamma_\textsc{k}}}
\newcommand{\E}{\ensuremath{\bm{E}}}
\newcommand{\B}{\ensuremath{\bm{B}}}

\newcommand{\mysub}[1]{\ensuremath{_{\mathrm{#1}}}}

\newcommand{\myerror}[2][NONE]{%
  \ifthenelse { \equal {#1} {NONE} } %
  {\ensuremath{\pm #2}}%
  {\ensuremath{_{-#1}^{#2}}}%
}

\newcommand{\hess}{H.E.S.S.\xspace}
\newcommand{\lhaaso}{LHAASO\xspace}
\newcommand{\magic}{MAGIC\xspace}

\newcommand{\grbLI}{GRB221009A\xspace}
\newcommand{\grbMI}{GRB190114C\xspace}
\newcommand{\grbHI}{GRB180728B\xspace}
\newcommand{\grbHII}{GRB190829A\xspace}

\definecolor{dg}{rgb}{0.0, 0.6, 0.1}
\definecolor{ed}{rgb}{1.0, 0.6, 0.1}
\makeatletter
\def\Andrew{\def\xx{dg}\@ifnextchar[{\@mwith}{\@mwithout}}
\def\Mitya{\def\xx{orange}\@ifnextchar[{\@mwith}{\@mwithout}}
\def\Felix{\def\xx{ed}\@ifnextchar[{\@mwith}{\@mwithout}}
\def\@mwith[#1]#2{\textcolor{\xx}{\sout{#1}#2}}
\def\@mwithout#1{\textcolor{\xx}{#1}}
\makeatother

\newcommand{\dias}{{Dublin Institute for Advanced Studies, School of Cosmic Physics, 31 Fitzwilliam Place, Dublin 2, Ireland}}

\newcommand{\mpik}{{Max-Planck-Institut f\"ur Kernphysik, Saupfercheckweg 1, 69117 Heidelberg, Germany}}
\newcommand{\yerevan}{{Yerevan State University, 1 Alek Manukyan St., Yerevan 0025, Armenia}}
\newcommand{\rikkyo}{{Graduate School of Artificial Intelligence and Science, Rikkyo University, Nishi-Ikebukuro 3-34-1, Toshima-ku, Tokyo 171-8501, Japan}}

\newcommand{\desy}{{DESY, Platanenallee 6, D-15738 Zeuthen, Germany}}

\newcommand{\Tau}{\mathrm{T}}
\usepackage{soul}
\shorttitle{TeV light curve of \grbLI}
\shortauthors{Khangulyan et al.}

\graphicspath{{./}{figures/}}

\begin{document}

\title{Naked forward shock seen in the TeV afterglow data of \grbLI}

\correspondingauthor{Dmitry Khangulyan}
\email{d.khangulyan@rikkyo.ac.jp}

\author[0000-0002-7576-7869]{Dmitry Khangulyan}
\affiliation{\rikkyo}

\author[0000-0003-1157-3915]{Felix Aharonian}
\affiliation{\dias}
\affiliation{\mpik}
\affiliation{\yerevan}

\author[0000-0001-9473-4758]{Andrew M. Taylor}
\affiliation{\desy}



\begin{abstract}
  We explore the implications of the light curve of the early TeV gamma-ray afterglow of GRB221009A reported by the
  LHAASO collaboration. We show that the reported offset of the reference time, \(T_*\), allows the determination of the relativistic jet activation time, which occurs approximately \(200\,\mathrm{s}\) after the GBM trigger time and closely
  precedes the moment at which GBM was saturated.  We find that while the LHAASO data do not exclude the homogeneous
  circumburst medium scenario, the progenitor wind scenario looks preferable, finding excellent agreement
  with the expected size of the stellar bubble. We conclude that the initial growth of the light curve is dominated by
  processes internal to the jet or by gamma-gamma attenuation on the photons emitted during the prompt phase. Namely,
  either the activation of the acceleration process or the decrease of internal gamma-gamma absorption can naturally
  explain the initial rapid flux increase. The subsequent slow flux growth phase observed up to \(T_*+18\,\mathrm{s}\) is
  explained by the build-up of the synchrotron radiation --- the target for inverse Compton scattering, which is
  also supported by a softer TeV spectrum measured during this period. The duration of this phase allows an almost
  parameter-independent determination of the jet's initial Lorentz factor, \(\Gamma_0\approx600\), and magnetic field
  strength, \(B'\sim0.3\,\mathrm{G}\). These values appear to match well those previously revealed through spectral
  modeling of the GRB emission.
\end{abstract}



\section{Introduction} \label{sec:intro}
Gamma-ray bursts (GRBs) result from gigantic explosions in the Universe occurring at redshifts on average  \(z\sim 2 -3\). 
These events are believed to be powered by ultra-relativistic outflows formed either by the collapse of massive stars or binary system mergers. Thanks to the bright, non-thermal emission generated in their outflows, GRBs are detected in a broad range of frequency bands. Synchrotron emission is believed to dominate from X- up to MeV gamma-ray energies, while the radiation in the GeV and TeV bands emerges from the inverse Compton (IC) scattering.
The detection of TeV gamma rays from GRBs is considered an important tool for constraining the physical conditions in the production region. Unfortunately, the attenuation by extragalactic background light (EBL), can severely hinder the detection of this component from GRBs at cosmological distances. So far, only a few GRBs have been detected in the TeV regime \citep{2019Natur.575..455M,2019Natur.575..464A,2021Sci...372.1081H},   
revealing that the GRB afterglow phase is characterized by bright TeV emission. 

The position of \grbLI at the trigger time, \(T_0\), serendipitously appeared in the \lhaaso field of view. 
Thanks to the extraordinarily high flux of TeV radiation and the superior sensitivity of the LHAASO detectors, the growth phase of the TeV emission associated with \grbLI was detected in the background-free observation regime. More than \(60,000\) very-high-energy (VHE) photons have been detected within the first \(3,000\unit{s}\) after the trigger, providing unprecedented TeV photon statistics allowing for the potential detection of $\geq 10 \%$   fluctuations on \(\sim1\unit{min}\) time scales. However, the TeV light curve demonstrated a very smooth evolution.   Together with the accurately measured time delay, this smooth behavior of the light curve suggests that the TeV emission originates from the forward shock, i.e., represents the early afterglow emission phase. Note that \grbMI was detected in the same early afterglow stage by \magic but with much lower statistics \citep{2019Natur.575..455M}. The two other TeV GRBs afterglows, \grbHI and \grbHII, were detected with \hess in late afterglow phase \citep{2019Natur.575..464A,2021Sci...372.1081H}.

While light curves covering early afterglow phases have already been obtained for many GRBs in the X-rays, MeV, and occasionally also GeV gamma-ray bands, the prompt phase emission strongly dominates in these bands at the early epochs making challenging the study of the onset of the afterglow emission with X-ray and MeV/GeV data only \citep[see, however,][]{2010A&A...510L...7G}. \lhaaso observations of \grbLI reveal no evidence for TeV emission at the prompt phase. Thus, the early TeV light curve provides us with a unique opportunity to study the early afterglow physics, i.e., the processes occurring at the forward shock,  without contamination from
the prompt emission. Thus, \grbLI offers the unique opportunity to observe the ``naked'' launching of a shock into the circumburst medium. In particular, the TeV light curve helps us to understand the initial phase's shock dynamics and the associated particle acceleration mechanism.

Several factors may significantly affect the early afterglow light curve, being, however, less important during the late afterglow phase. For example, spectral modeling of afterglow emission favors a Gauss-strength magnetic field in the production region  \citep[e.g.,][]{2019ApJ...880L..27D}, which requires that the magnetic field is significantly amplified. Processes related to the magnetic field evolution are believed to be key factors leading to the particle acceleration at relativistic shocks \citep[see][for a review]{10.1063/1.3701361}, and the time required for their activation may lead to a delay in the onset of TeV particle acceleration. Also, at the initial stage, the blast wave propagates through an inhomogeneous environment, which includes the SN shell, stellar wind, its termination shock, and the stellar bubble   (i.e., a layer of shocked stellar wind and compressed interstellar medium). As the blast wave's propagation proceeds in the ultrarelativistic regime, this inhomogeneity may cause an apparent delay concerning the trigger time. In this study, we focus on effects related to the shock dynamics at the early afterglow stage and investigate the implications of the \lhaaso observation of \grbLI on the properties of the circumburst medium. 

\section{Shock dynamics}\label{sec:shock_dynamics}
The self-similar solution by \citet[hereafter BM76]{1976PhFl...19.1130B} provides a good description for the dynamics of the forward shock only when the energy of the swept-up shell, \(M\), is comparable to the explosion energy, \(E_0\lesssim \Gamma_0^2M c^2\), i.e., when
\be\label{eq:BM_condition}
M\gtrsim 6\times10^{-5}\mathrm{M}_\odot\qty(\frac{E_0}{10^{55}\unit{erg}})\qty(\frac{\Gamma_0}{300})^{-2}\,.
\ee
An extrapolation of the BM76 solution to the very beginning of the explosion is, however, incorrect. 
During the initial explosion phase, the propagation of the blast wave is little affected by the circumburst medium (refered to as the ``coasting phase'', an analogue of the ``ejecta-dominate phase'' for supernovae explosions). Thus the blast wave moves with nearly constant Lorentz factor, and lags behind the photon front which propagates out from the explosion origin differently than that predicted from the BM76 solution \citep{2007ApJ...655..973K}. This lag leads to an offset of the self-similar solution reference time, \(T_*\), with respect to the trigger time, \(T_0\), even if the trigger time accurately determines the moment when the GRB jet starts propagating from its origin. We note additionally that the GRB precursor can be triggered on, in which case the reported trigger time may refer (at least for some specific cases) to a time prior to the jet activation.

Since the blast wave propagates in the ultra-relativistic regime, one needs to account for relativistic effects and kinematic delays of the signal. Letting \(\tau\) be the time elapsed since the explosion in the progenitor reference frame; \(\tau'\) is the time in the blast co-moving frame; and \(t\) is the detection time (which is measured relative to the trigger time) of (hypothetical) photons emitted at the shock front at time \(\tau\). If the blast wave Lorentz factor is \(\Gamma\), then \(\dd{\tau} = \Gamma\dd{\tau'}\) due to the relativistic time dilation. If two photons were emitted at the blast wave front at \(\tau\) and \(\tau+\dd{\tau}\), when their detection time is separated by the observer time interval of \(\dd{t}=\dd{\tau}/(2\Gamma^2)\), as the blast wave fronts moves with \(v\approx c\qty(1-\nicefrac{1}{(2\Gamma^2)})\). Since the blast wave Lorentz factor changes as the wave propagates, then
\be\label{eq:tau_gen}
t_2-t_1 \approx \int\limits_{\Tau_1}^{\Tau_2}\frac{\dd{\tau}}{2\Gamma^2(\tau)}\approx\int\limits_{R_1}^{R_2}\frac{\dd{r}}{2c\Gamma^2(r)}\,.
\ee
If a photon is emitted from the blast wave position at the moment when the wave reaches a distance \(R\) from the explosion origin (the distance is measured in the progenitor frame),  the observer detects this photon delayed with respect to the trigger time by
\be\label{eq:tau_trig}
t \approx\int\limits_{0}^{R}\frac{\dd{r}}{2c\Gamma^2(r)}\,.
\ee
During the coasting phase, 
the bulk Lorentz factor of the forward shock is roughly a constant, \(\Gamma_0\). Thus, Eq.~\eqref{eq:tau_trig} is reduced to \(t\approx R/(2c\Gamma_0^2)\).  As the initial Lorentz factor is expected to be large, \(\Gamma_0\gg 10^2\), the blast wave overtakes the supernovae shell already on a very short time interval for the observer:
\be
t\mysub{shell}\sim 20 \qty(\frac{R\mysub{shell}}{10^{12}\unit{cm}})\qty(\frac{\Gamma_0}{300})^{-2}\unit{ms}\,.
\ee
Since the \lhaaso photon statistics for \grbLI corresponds to \(0.02\unit{photon\,ms^{-1}}\), resolving such millisecond times-scales remains challenging even for present day gamma-ray detectors.

\subsection{Interaction with stellar wind}
\begin{figure}
  \plotone{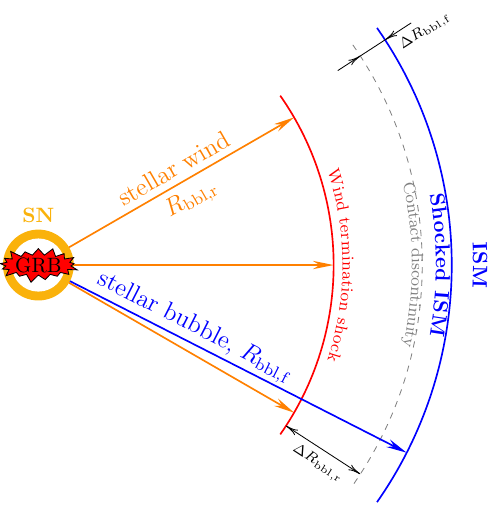}
  \caption{Structure of the circumburst environment. The GRB jet initially propagates through the fast stellar wind, then susbsequently interacts with the stellar bubble (shocked stellar wind and compressed ISM layer). After that the jet finally reaches the regular ISM. \label{fig:env}}
\end{figure}

\begin{figure}
  \includegraphics[height=18cm]{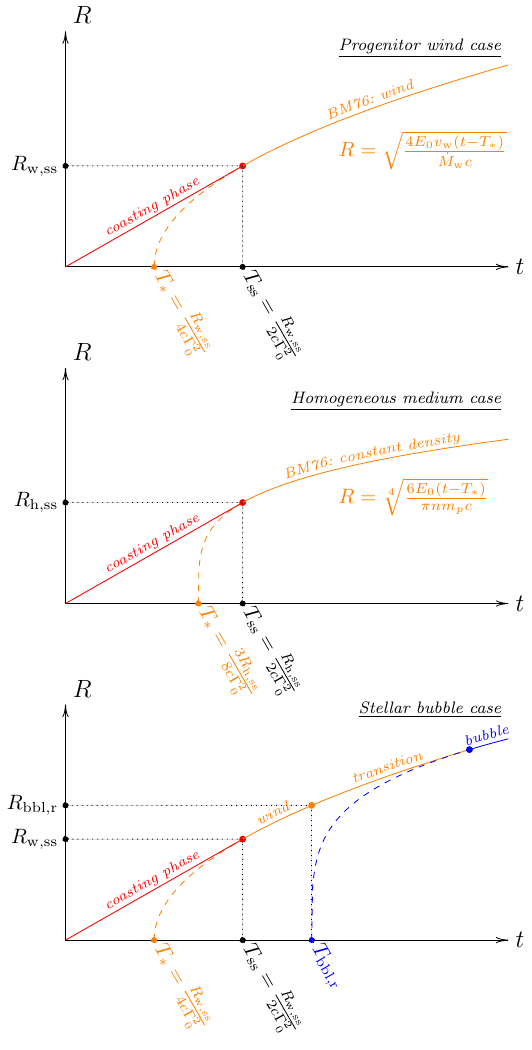}
  \caption{Transition from coasting phase to BM76 self-similar solution results in an offset of the reference time in respect to the trigger time. Top panel: GRB jet interacting with the progenitor wind; middle panel: GRB jet interacting with homogeneous medium; bottom panel: GRB jet reaches the stellar bubble. \label{fig:coasting_BM}}
\end{figure}

As illustrated in Fig.~\ref{fig:env}, after emerging through the SN shell, the blast wave starts interacting with the stellar wind. The typical stellar wind speed is \(v\mysub{w}\approx2\times 10^{3}\unit{km\,s^{-1}}\) and the mass loss rates of a massive star can be quite high, say \(M\mysub{w}\sim 10^{-7}\mathrm{M}_\odot\unit{yr^{-1}}\). Thus, the blast wave converges to the self-similar solution when it reaches a distance of
\be\label{eq:wind_ss}
\begin{split}
R\mysub{w,ss}&\sim\frac{E_0v\mysub{w}}{\Gamma_0^2 c^2\dot{M}\mysub{w}}\\
             &\sim 1 \qty(\frac{E_0}{10^{55}\unit{erg}})\qty(\frac{\Gamma_0}{300})^{-2}\times\\
             &\qty(\frac{v\mysub{w}}{2\times10^3\unit{km\,s^{-1}}})\qty(\frac{\dot{M}\mysub{w}}{10^{-7}\mathrm{M}_\odot})^{-1}\unit{pc}\,.
\end{split}                  
\ee
In principle, this distance may exceed the inner size of the stellar bubble, \(R\mysub{bbl,r}\). However, for the present discussion we assume that it is smaller.

For the range of distances \(R\mysub{w,ss}<R<R\mysub{bbl,r}\), the blast wave propagates in the self-similar regime within the circumburst density \(n\mysub{c}\propto R^{-2}\). In this case, the blast wave Lorentz factor is
\be
\Gamma\approx \sqrt{\frac{E_0 v\mysub{w}}{R \dot{M}\mysub{w}c^2}}\,.
\ee
Thus, according to Eq.~\eqref{eq:tau_gen}, the blast wave front lags behind the trigger photon front by
\be\label{eq:tau_wind}
t = \left\{
  \begin{matrix}
    \frac{R}{(2c\Gamma_0^2)}&\qq{for}&R<R\mysub{w,ss}\\
    \frac{R\mysub{w,ss}}{(2c\Gamma_0^2)}+\frac{\qty(R^2- R\mysub{w,ss}^2)\dot{M}\mysub{w}c}{4E_0 v\mysub{w}}&\qq{for}&R\mysub{w,ss}<R<R\mysub{bbl,r}\,.\\
  \end{matrix}
\right.
\ee
Normalizing the parameters in Eq.~\eqref{eq:tau_wind} one obtains
\be\label{eq:tau_wind_norm}
\begin{split}
  \frac{R}{(2c\Gamma_0^2)}&=6\times10^2\qty(\frac{R}{1\unit{pc}})\qty(\frac{\Gamma_0}{300})^{-2}\unit{s}\\
  \frac{\qty(R^2- R\mysub{w,ss}^2)\dot{M}\mysub{w}c}{4E_0 v\mysub{w}}&=2.2\times10^2\qty(\frac{R^2- R\mysub{w,ss}^2}{1\unit{pc^2}})\times\\
\end{split}
\ee
\[
    \qty(\frac{\dot{M}\mysub{w}}{10^{-7}\mathrm{M}_\odot\unit{yr^{-1}}})\qty(\frac{E_0}{10^{55}\unit{erg}})^{-1}\qty(\frac{v\mysub{w}}{2\times10^3\unit{km\,s^{-1}}})^{-1}\unit{s}\,.
\]

When the blast wave propagates in the self-similar regime, the corresponding reference time differs from the trigger time. Using Eq.~\eqref{eq:tau_wind} one can obtain this offset of the reference time with respect to the trigger time (see in Fig.~\ref{fig:coasting_BM}) as:
\be\label{eq:wind_shift}
\begin{split}
  T_{*}&\approx \frac{R\mysub{w,ss}}{2c\Gamma_0^2} - \frac{R\mysub{w,ss}^2\dot{M}\mysub{w}c}{4E_0 v\mysub{w}}\approx\frac{R\mysub{w,ss}}{4c\Gamma_0^2}\\
            &\approx 3\times10^2\qty(\frac{R\mysub{w,ss}}{1\unit{pc}})\qty(\frac{\Gamma_0}{300})^{-2}\unit{s}\,,
\end{split}
\ee
where we also accounted for Eq.~\eqref{eq:wind_ss}.  Thus, the blast wave radius is
\be\label{eq:radius_wind}
R\approx\left\{
  \begin{matrix}
    2c\Gamma_0^2t&\qq{for}&t< \frac{R\mysub{w,ss}}{2c\Gamma_0^2}\\
    \sqrt{\frac{4E_0v\mysub{w}\qty(t-T_*)}{\dot{M}\mysub{w}c}}&\qq{for}&t> \frac{R\mysub{w,ss}}{2c\Gamma_0^2}\,,
  \end{matrix}
\right.
\ee
where
\be
2c\Gamma_0^2t\approx 5\times10^{17}\qty(\frac{\Gamma_0}{300})^2\frac{t}{100\unit{s}}\unit{cm}
\ee
and
\be
\sqrt{\frac{4E_0v\mysub{w}\qty(t-T_*)}{\dot{M}\mysub{w}c}}\approx2\times10^{18}\qty(\frac{E_0}{10^{55}\unit{erg}})^{\nicefrac{1}{2}}\times
\ee
\[
\qty(\frac{v\mysub{w}}{2\times10^{3}\unit{km\,s^{-1}}})^{\nicefrac{1}{2}}\qty(\frac{\dot{M}\mysub{w}}{10^{-7}\mathrm{M}_\odot\unit{yr^{-1}}})^{\nicefrac{-1}{2}}\qty(\frac{t-T_*}{100\unit{s}})^{\nicefrac{1}{2}}\unit{cm}\,.
\]
We note that the offset of the reference time appears only for the self-similar regime of the shock wave propagation.

\subsection{Homogeneous circumburst medium}
Another standard case, which is typically considered in the literature, is the interaction of the GRB jet with a homogeneous circumburst medium, which is often dubbed as ISM. 
According Eq.~\eqref{eq:BM_condition}, in the case of homogeneous circumburst medium,  the transition to the self-similar regime occurs at
\be
\begin{split}
  R\mysub{h,ss}&\approx \sqrt[3]{\frac{3}{4\pi}\frac{E_0}{\Gamma_0^2m_pn c^2}}\\
               &\approx0.1\qty(\frac{E_0}{10^{55}\unit{erg}})^{\nicefrac{1}{3}}\qty(\frac{n}{1\unit{cm^{-3}}})^{\nicefrac{-1}{3}}\qty(\frac{\Gamma_0}{300})^{\nicefrac{-2}{3}}\unit{pc}\,.
\end{split}
\ee
Here \(m_p\) is the proton mass and \(n\) is the number density of homogeneous medium. For an explosion in a constant density environment, the blast wave Lorentz factor, for \(R>R\mysub{h,ss}\), is
\be
\Gamma\approx \sqrt{\frac{E_0}{\left(\nicefrac{4\pi}{3}\right) R^3 nm_pc^2}}\,.
\ee
In this case, the self-similar solution implies that the blast wave front lags behind the trigger photon front by
\be\label{eq:tau_ism}
t = \left\{
  \begin{matrix}
    \frac{R}{(2c\Gamma_0^2)}&\qq{for}&R<R\mysub{h,ss}\\
    \frac{R\mysub{h,ss}}{(2c\Gamma_0^2)}+\frac{\pi}{6}\frac{nm_pc\qty(R^4-R\mysub{h,ss}^4)}{E_0}&\qq{for}&R\mysub{h,ss}<R\,.\\
  \end{matrix}
\right.
\ee
Normalizing the parameters in Eq.~\eqref{eq:tau_ism} one obtains
\be\label{eq:tau_ism_norm}
\frac{\pi}{6}\frac{nm_pcR^4}{E_0}\approx 20\qty(\frac{E_0}{10^{55}\unit{erg}})^{-1}\qty(\frac{n}{1\unit{cm^{-3}}})\qty(\frac{R\mysub{bbl}}{0.1\unit{pc}})^{4}\unit{s}\,.
\ee
As shown in Fig.~\ref{fig:coasting_BM}, the difference between the trigger time and the reference time is
\be\label{eq:ism_shift}
\begin{split}
  T_*&\approx\frac{R\mysub{h,ss}}{2c\Gamma_0^2} - \frac{\pi}{6}\frac{nm_pcR\mysub{h,ss}^4}{E_0}=\frac{3R\mysub{h,ss}}{8c\Gamma_0^2}\,,\\
         &\approx40 \qty(\frac{R\mysub{h,ss}}{0.1\unit{pc}})\qty(\frac{\Gamma_0}{300})^{-2}\unit{s}\,,
\end{split}
\ee
and the shock radius is
\be
R\approx\left\{
  \begin{matrix}
    2c\Gamma_0^2t&\qq{for}&t<\frac{R\mysub{h,ss}}{2c\Gamma_0^2}\\
    \sqrt[4]{\frac{6E_0\qty(t-T_*)}{\pi nm_pc}}&\qq{for}&t>\frac{R\mysub{h,ss}}{2c\Gamma_0^2}\,,\\
  \end{matrix}
\right.
\ee
where
\be
\sqrt[4]{\frac{6E_0\qty(t-T_*)}{\pi nm_pc}}=4\times10^{17}\qty(\frac{E_0}{10^{55}\unit{erg}})^{\nicefrac{1}{4}}\times
\ee
\[
  \qty(\frac{n}{1\unit{cm^{-3}}})^{\nicefrac{-1}{4}}\qty(\frac{t-T_*}{100\unit{s}})^{\nicefrac{1}{4}}\unit{cm}\,.
\]

\subsection{Stellar bubble}\label{sec:bbl}

As shown in Fig.~\ref{fig:env}, before reaching the ISM, the shock should additionally propagate through the stellar bubble, which consists of two layers: the shocked stellar wind and the shocked ISM. The mass of the shocked stellar wind depends on the mass loss rate of the stellar wind and the star life time, \(\tau\mysub{star}\):
\be\label{eq:mass_shocked_wind}
\dot{M}\mysub{w} \tau\mysub{star}= 0.1\mathrm{M}_\odot\qty(\frac{\dot{M}\mysub{w}}{10^{-7}\mathrm{M}_\odot\unit{yr^{-1}}}) \qty(\frac{\tau\mysub{star}}{10^6\unit{yr}})\,,
\ee
which almost unavoidably appears larger than the amount of external gas required for the transition into the self-similar regime given by Eq.~(\ref{eq:BM_condition}). Thus, even if the blast wave traveled in the wind zone with constant Lorentz factor, it should start slowing down within the shocked stellar wind, well before reaching the shocked ISM. It is also possible to estimate the blast wave Lorentz factor at the moment of reaching the contact discontinuity at \(R\mysub{bbl,r}+\Delta R\mysub{bbl,r}\):
\be
\begin{split}
  \Gamma&\approx \frac{E_0}{\dot{M}\mysub{w}\tau\mysub{star}c^2}\\
        &\approx 60 \qty(\frac{E_0}{10^{55}\erg})\qty(\frac{\dot{M}\mysub{w}}{10^{-7}\mathrm{M}_\odot\unit{yr^{-1}}})^{-1} \qty(\frac{\tau\mysub{star}}{10^6\unit{yr}})^{-1}\,,
\end{split}
\ee
i.e., although the shock is still expected to be relativistic, it gets decelerated very significantly from its initial Lorentz factor, \(\Gamma_0\gg10^2\).

The outer radius of the stellar bubble can be estimated using the self-similar solution for non-relativistic gas:
\be\label{eq:stellar_bubble}
\begin{split}
  R\mysub{bbl,f}&\approx \sqrt[5]{\frac{L\mysub{w}\tau\mysub{star}^3}{m_pn\mysub{ism}}}\\
              &\sim 20\qty(\frac{L\mysub{w}}{10^{35}\unit{erg\,s^{-1}}})^{\nicefrac{1}{5}}\qty(\frac{\tau\mysub{star}}{10^{6}\unit{yr}})^{\nicefrac{3}{5}}\qty(\frac{n\mysub{ism}}{1\unit{cm^{-3}}})^{\nicefrac{-1}{5}}\unit{pc}\,.
\end{split}
\ee
Here \(L\mysub{w}=\dot{M}\mysub{w}v\mysub{w}^2/2\) is the kinetic luminosity of the stellar wind 
and \(n\mysub{ism}\) is the ISM density.

The shocked ISM layer density can be estimated from the shock compression ratio. For a strong non-relativistic shock, for a gas with polytropic gas index of $5/3$, a compression ratio of \(4\) is expected \citep[see, e.g.][]{Landau_FM_1987}. The layer thickness can be obtained from the mass conservation, which for a spherically symmetric configuration yields a value of approximately
\be
\Delta R\mysub{bbl,f}\approx0.1R\mysub{bbl,f}\,
\ee
i.e., it represents a quite thin compressed layer. The mass of the ISM gas swept-up by the stellar wind is expected to be very significant:
\be
M\mysub{bbl}\approx10^{2}\mathrm{M}_\odot \qty(\frac{n\mysub{ism}}{1\unit{cm^{-3}}}) \qty(\frac{R\mysub{bbl}}{10\unit{pc}})^{3}\,.
\ee

Expansion of this heavy external layer is supported by the inner layer that consists of stellar wind that passes through the stellar bubble reverse shock at \(R\mysub{bbl,r}\). This shock terminates the stellar wind and creates a layer of hot gas of approximately constant density. The density of the layer is determined by the jump conditions at a strong non-relativistic shock:
\be\label{eq:bbl_r_density}
\begin{split}
  n\mysub{bbl,r}&\approx \frac{\dot{M}\mysub{w}}{\pi R\mysub{bbl,r}^2v\mysub{w}m_p}\\
  &\sim10^{-3} \qty(\frac{\dot{M}\mysub{w}}{10^{-7}\mathrm{M}_\odot\unit{yr^{-1}}})\times
\end{split}
\ee
\[
  \qty(\frac{v\mysub{w}}{2\times10^{3}\unit{km\,s^{-1}}})^{-1}\qty(\frac{R\mysub{bbl,r}}{1\unit{pc}})^{-2}\unit{cm^{-3}}\,.
\]
As shown above, the contact discontinuity is located at approximately \(R\mysub{bbl,f}\), thus we can use the conservation of the mass ejected by the wind to obtain the inner radius of the stellar bubble, i.e., the radius of the wind termination shock:
\be
\frac{4\pi}{3} n\mysub{bbl,r}m_p\qty(R\mysub{bbl,f}^3-R\mysub{bbl,r}^3) = \dot{M}\mysub{w}\tau\mysub{star}\,.
\ee
Solving this equation one obtains the approximate solution
\be
\begin{split}
  \frac{R\mysub{bbl,r}}{R\mysub{bbl,f}}&\approx \sqrt{\frac{4R\mysub{bbl,f}}{3\tau\mysub{star}v\mysub{w}}}\\
  &\approx 0.1\qty(\frac{R\mysub{bbl,f}}{30\unit{pc}})^{\nicefrac{1}{2}}\qty(\frac{\tau\mysub{star}}{10^{6}\unit{yr}})^{\nicefrac{-1}{2}}\qty(\frac{v\mysub{w}}{2\times10^{3}\unit{km\,s^{-1}}})^{\nicefrac{-1}{2}}
\end{split}
\ee
Thus, one should expect the following structure of the stellar bubble around a massive star: the forward shock of typical radius of \(\sim20\unit{pc}\), a contact discontinuity at almost the same distance, and the reverse shock at a few parsec from the explosion origin.

When the relativistic blast wave reaches the reverse shock it starts interacting with the homogeneous medium with the typical density given by Eq.~\eqref{eq:bbl_r_density}.
Subsequently, once the explosion energy has been transferred to the shocked stellar wind, the further propagation of the shock proceeds in the self-similar regime expected for a homogeneous medium, \(n\propto \mathrm{const}\). In Appendix \ref{app:bubble} we provide analytic expression for the solution describing the shock propagation in the stellar bubble. 

In this case, the delay of the blast wave propagating though the shocked stellar wind behind the trigger photon front for \(R>R\mysub{bbl}\) is given by 
\be
\begin{split}
t\approx& t\qty(R\mysub{bbl,r})+\frac{\dot{M}\mysub{w}cR\mysub{bbl,r}\qty(R-R\mysub{bbl,r})}{2E_0v\mysub{w}}+ \\
        &\frac{\pi}{6}\frac{n\mysub{bbl,r}m_pc\qty(R^4-R\mysub{bbl,r}^4)}{E_0}-\\
        &\frac{4\pi}{6}\frac{n\mysub{bbl,r}m_pc\qty(R-R\mysub{bbl,r})R\mysub{bbl,r}^3}{E_0}\,,
\end{split}
\ee
where \(t\qty(R\mysub{bbl,r})\) is the delay accumulated in the unshocked wind:
\be\label{eq:delay_wind}
t\qty(R\mysub{bbl,r})=\left\{
  \begin{matrix}
    \frac{R\mysub{bbl,r}}{(2c\Gamma_0^2)}&\qq{if}&R\mysub{bbl,r}<R\mysub{w,ss}\\
    \frac{R\mysub{w,ss}}{(4c\Gamma_0^2)}+\frac{R\mysub{bbl,r}^2\dot{M}\mysub{w}c}{4E_0 v\mysub{w}}&\qq{if}&R\mysub{bbl,r}>R\mysub{w,ss}\,.\\
  \end{matrix}
\right.
\ee
The size of the stellar bubble reverse shock determines the reference time for the self-similar solution (which should emerge for \(R\gg R\mysub{bbl,r}\)):
\be\label{eq:bbl_shift}
\begin{split}
  T\mysub{bbl,*}&\approx t\qty(R\mysub{bbl,r}) - \frac{\dot{M}\mysub{w}cR\mysub{bbl,r}^2}{2E_0v\mysub{w}} + \frac{\pi}{2}\frac{n\mysub{bbl,r}m_pcR\mysub{bbl,r}^4}{E_0}\\
 &\approx\frac{R\mysub{w,ss}}{(4c\Gamma_0^2)}+\frac{R\mysub{bbl,r}^2\dot{M}\mysub{w}c}{4E_0 v\mysub{w}} = t\qty(R\mysub{bbl,r}) \,,
\end{split}
\ee
where we also used Eq.~\eqref{eq:bbl_r_density}.
The normalized terms in Eqs.~\eqref{eq:delay_wind} and \eqref{eq:bbl_shift} are given by Eq.~\eqref{eq:tau_wind_norm} and \eqref{eq:tau_ism_norm}. 
We note here, that under our adopted simplified description the reference time for the shock propagating through the stellar bubble coincides with the time when the shock reaches the inner boundary of the stellar bubble. Thus, one can use the light curve to determine this specific moment of time. 

Thus, the dependence of the blast wave radius (which determines the Lorentz factor) for \(R<R\mysub{bbl,r}\) follows Eq.~\eqref{eq:radius_wind}, and for \(R\gg R\mysub{bbl,r}\) 
\be\label{eq:radius_bbl}
\begin{split}
  R&\approx\sqrt[4]{\frac{6E_0\qty(t-T\mysub{bbl,*})}{\pi n\mysub{bbl,r}m_pc}}\,,\\
   &\approx 2\times10^{18}\qty(\frac{E_0}{10^{55}\unit{erg}})^{\nicefrac{1}{4}}\qty(\frac{n\mysub{bbl,r}}{10^{-3}\unit{cm^{-3}}})^{\nicefrac{-1}{4}}\times\\
  &\quad\quad\qty(\frac{t-T_*}{100\unit{s}})^{\nicefrac{1}{4}}\unit{cm}\,.
\end{split}
\ee
We note that in this case two different reference times appear in the solution: in the range \(R\mysub{w,ss}<R<R\mysub{bbl,r}\) the reference time is given by Eq.~\eqref{eq:wind_shift} and for \(R\gg R\mysub{bbl,r}\) the reference time is given by Eq.~\eqref{eq:bbl_shift}. 
\section{Light curve}
In addition to the change of the bulk Lorentz factor discussed in the previous section, there are several basic processes that operate simultaneously in the production region which can also imprint themselves onto the afterglow emission lightcurve. These include particle acceleration, formation of the target, gamma-gamma attenuation, and injection of energy into the production region. We next discuss each of these in turn.

\subsection{Acceleration}
The particle acceleration process operates in the shock co-moving frame (note that here, for the sake of simplicity, we do not distinguish the downstream and shock reference frames) and the most basic acceleration time-scale depends on the magnetic field strength \(B'\) in the downstream:
\be\label{eq:acc}
\tau'\mysub{acc}\sim \frac{\eta E'}{eB'c}\approx 0.1 \eta\mysub{acc} \qty(\frac{E'}{1\unit{TeV}})\qty(\frac{B'}{1\unit{G}})^{-1}\unit{s}\,.
\ee
Here \(\eta\mysub{acc}=B'/{\cal E}\) is a phenomenological factor determining the acceleration efficiency where \({\cal E}\) denotes the accelerating electric field, and \(E'\) is the particle energy in the comoving frame. The acceleration time converts to a detection delay of
\be
t\mysub{acc}\approx\frac{\tau'\mysub{acc}}{2\Gamma}\approx0.2 \eta\mysub{acc} \qty(\frac{E'}{1\unit{TeV}})\qty(\frac{B'}{1\unit{G}})^{-1}\qty(\frac{\Gamma}{300})^{-1}\unit{ms}\,,
\ee
i.e., it cannot cause any considerable delay unless the magnetic field is very weak, \(B'\sim\unit{mG}\), and the blast wave Lorentz factor is small, \(\Gamma<10\). We note, however, that this estimate does not account for the time required for activation of the acceleration process, in particular for the magnetic field amplification. For example, it was revealed with \hess observations of RS Ophiuchi  that this time scale can be an important factor delaying the onset of TeV emission from shocks accelerating in the non-relativistic regime \citep{2022Sci...376...77H}.

\subsection{Target development}\label{sec:target}
If one considers SSC as the radiation process dominating gamma-ray emission in the VHE band, the typical time-scale to create the target is determined by the cooling time of electrons that provide the synchrotron photons. To generate TeV emission in the observer frame (\(\hbar\omega\sim1\unit{TeV}\)), the comoving energy of emitting electrons can be roughly estimated as \(\hbar \omega /\Gamma\). These electrons up-scatter target photons with energy \(\hbar \omega\mysub{ph}' < \Gamma m_e^2c^4/(\hbar \omega)\approx 100\qty(\hbar\omega/\unit{TeV})^{-1}\qty(\Gamma/300)\unit{eV}\)  the most efficiently \citep[see, e.g.,][references therein]{2023arXiv230712467K}. Electrons emitting these photons have energy 
\be
E'_e\lesssim40\qty(\frac{B'}{1\unit{G}})^{\nicefrac{-1}{2}}\qty(\frac{\hbar\omega}{1\unit{TeV}})^{\nicefrac{-1}{2}}\qty(\frac{\Gamma}{300})^{\nicefrac{1}{2}}\unit{GeV}\,.
\ee
Thus the typical time scale for developing the target field is
\be\label{eq:target}
\tau'\mysub{ph}\gtrsim 10^{4} \qty(\frac{B'}{1\unit{G}})^{\nicefrac{-3}{2}}\qty(\frac{\hbar\omega}{1\unit{TeV}})^{\nicefrac{1}{2}}\qty(\frac{\Gamma}{300})^{\nicefrac{-1}{2}}\unit{s}\,.
\ee
This time scale corresponds to a detection delay of 
\be\label{eq:tau_target}
t\mysub{ph}\approx\frac{\tau'\mysub{ph}}{2\Gamma}\approx20 \qty(\frac{B'}{1\unit{G}})^{\nicefrac{-3}{2}}\qty(\frac{\hbar\omega}{1\unit{TeV}})^{\nicefrac{1}{2}}\qty(\frac{\Gamma}{300})^{\nicefrac{-3}{2}}\unit{s}\,.
\ee
This delay is non-negligible even if the magnetic field in the production region has Gauss strength, and for weaker values the delay due to this process would become even larger.

If the target development is indeed responsible for the formation of part ``\emph{a}'' of the light curve, one would expect for this to be accompanied by a corresponding spectral transformation of the IC spectrum. This transformation is caused by the change of the slope of the target photon field, which transitions from a slow cooling to a fast cooling spectra. If the power-law index of the cooled electron  distribution is \(\alpha\), then the photon index of synchrotron slow cooling spectrum is \(\alpha/2\), which softens to \((\alpha+1)/2\) in the fast cooling regime. According to \cite{2023arXiv230712467K}, in the former case the IC spectrum should be \(\alpha/2 +1\), which in the latter case becomes \(\alpha/2+1/2\), i.e., one expects a hardening of IC once the photon target development is completed.

\subsection{Absorption}\label{sec:absorption}
High-energy photons can effectively interact with low-energy target photons to create electron-positron pairs. The maximum of the cross-section, \(\sigma\mysub{e^+e^-}\approx0.26\sigma\mysub{T}\) (here \(\sigma\mysub{T}\) is the Thomson cross section), is achieved when TeV gamma-rays interact with 
\be
\ve\mysub{ph}\approx\frac{2.8m_e^2c^4}{\qty(1-\cos\theta)\hbar\omega}\approx 0.1\qty(\frac{\Gamma}{300})^2\qty(\frac{\hbar \omega}{1\unit{TeV}})^{-1}\unit{MeV}
\ee
target photons. Here we consider an interaction with a specific scattering angle, \(\theta\), and we have taken into account that for a relativistically moving shell the photons have a highly anisotropic angular distribution, thus \(\qty(1-\cos\theta)\approx 1/\qty(2\Gamma^2)\). 
Using the standard expression for the optical depth, one obtains that the gamma-gamma absorption is important if
\be
\frac{0.26\sigma\mysub{T}\qty(1-\cos\theta)L\mysub{ph}}{4\pi R c \ve\mysub{ph}}> 1\,.
\ee
Here \(L\mysub{ph}\) should be understood as the luminosity of the source responsible for target photons in the range of frequencies approximately from \(\approx\ve\mysub{ph}/1.5\) to \(\approx1.5\ve\mysub{ph}\), i.e spanning a factor of \(\approx2\) in frequency. For attenuation of TeV gamma-ray photons, emitted from a shell moving with bulk Lorentz factor of \(\simeq300\),  MeV frequency band plays the most important role.

Thus, we obtain a lower limit for the luminosity of the target photon source above which gamma-gamma absorption becomes important:
\be\label{eq:atten_lum}
\begin{split}
  L\mysub{ph}\qty(\ve\mysub{ph})&>10^3\Gamma^4\frac{Rm_e^2c^5}{\sigma\mysub{T}\hbar\omega}\\
                                &\gtrsim 10^{52}\qty(\frac{\Gamma}{300})^4\qty(\frac{R}{10^{17}\unit{cm}})\qty(\frac{\hbar\omega}{1\unit{TeV}})^{-1}\ergs\,.
\end{split}
\ee

\subsection{Energy supply}
Another key factor determining the light curve dependence is the rate at which energy is supplied to the production region. 
While the bulk Lorentz factor is sufficiently high, \(\Gamma>\theta\mysub{jet}^{-1}\) (where \(\theta\mysub{jet}\) is GRB jet half-opening angle), the observer can detect photons emitted from only a small fraction of the blast wave, i.e., one can use a spherically symmetric approximation for the blast wave. In this case the luminosity of the shock is simply
\be\label{eq:shock_lum0}
L\mysub{iso}\approx 4\pi R^2\eta \Gamma^4m_pn\mysub{c}c^3\,,
\ee
where \(\eta\) is the radiation efficiency, determined as a ratio of energy emitted the shocked gas to the kinetic energy of the gas entering in the production region.
Although Eq.~\eqref{eq:shock_lum0} is very basic, it nevertheless allows one to derive insightful conclusions. If the blast wave interacts with the stellar wind, \(n\mysub{c}\propto R^{-2}\), then the luminosity is determined by the dependence on time of the Lorentz factor and the radiation efficiency:
\be\label{eq:shock_lum}
L\mysub{w,iso}\approx \frac{\eta \Gamma^4\dot{M}\mysub{w}c^3}{v\mysub{w}}\,,
\ee

During the coasting phase the Lorentz factor is constant, thus the luminosity is determined by the dependence of \(\eta\):
\be
L\mysub{w,iso} \approx 7\times10^{51}\eta\qty(\frac{\dot{M}\mysub{w}}{10^{-7}\mathrm{M}_\odot\unit{yr^{-1}}})\times
\ee
\[
  \qty(\frac{v\mysub{w}}{2\times10^3\unit{km\,s^{-1}}})^{-1}\qty(\frac{\Gamma_0}{300})^4\unit{erg\,s^{-1}}\,.
\]
When the blast expansion enters into the self-similar regime, one needs to account for the Lorentz factor dependence of \(t\):
\be
\Gamma = \sqrt[4]{\frac{E_0v\mysub{w}}{4\dot{M}\mysub{w}c^3\qty(t-T_*)}}\,.
\ee
Thus, the luminosity is simply
\be
L\mysub{w,iso}=\frac{\eta}4\frac{E_0}{\qty(t-T_*)}\,.
\ee
We note that here, for the sake of simplicity, we assume an \(\omega^{-2}\) emission spectrum.
When the jet break occurs, i.e., the blast wave Lorentz factor drops below the inverse jet half-opening angle, \(\Gamma<\theta\mysub{j}^{-1}\), then one needs to introduce an additional factor \(\qty(\Gamma\theta\mysub{j})^2\), which leads to a break by \(0.5\) in the light curve:
\be
L\mysub{w,jb}\propto\qty(t-T_*)^{\nicefrac{-3}{2}}\,.
\ee

If the blast wave interacts with the homogeneous medium the situation is quite different. If the Lorentz factor is initially constant, the luminosity is
\be\label{eq:coasting_ism_iso}
L\mysub{h,iso}\approx 16\pi\eta \Gamma_0^8t^2m_pnc^5\,.
\ee
where we have accounted for the relation between the radius of the blast wave and time lag: \(R=2c\Gamma_0^2t\).

Once the expansion rate approaches the self-similar solution, one needs to account for the change of the Lorentz factor and for the dependence of \(R\) on \(t\):
\be
\Gamma \approx \frac12\sqrt[8]{\frac{3E_0}{8\pi m_pn c^5\qty(t-T_*)^3}}\,.
\ee
Thus, one obtains 
\be\label{eq:shock_ism_iso}
L\mysub{h,iso}\approx \frac{3\eta}{8}\frac{E_0}{\qty(t-T_*)}\,,
\ee
where one also accounted for \(R\approx \sqrt[3]{\frac{E_0}{\left(\nicefrac{4\pi}{3}\right)m_pn c^2\Gamma^2}}\).
After the jet break, one needs to account for an additional factor \(\qty(\theta\mysub{j}\Gamma)^2\propto(t-T_*)^{\nicefrac{-3}{4}}\), thus one obtains
\be\label{eq:shock_ism_jb}
L\mysub{h,jb}\propto\qty(t-T_*)^{\nicefrac{-7}{4}}\,.
\ee

It is also important to emphasize that the offset of the reference time appears only in the self-similar phase of the blast wave propagation: thus in  Eq.~\eqref{eq:coasting_ism_iso} \(t\) is measured from the moment of the jet launch, but in Eqs.~\eqref{eq:shock_ism_iso} and \eqref{eq:shock_ism_jb} an offset of the reference time, \(T_*\), appears.
\section{\lhaaso observations} \label{sec:observations}

The obtained light curve is consistent with a broken power-law, whose reference time is \(T_*=T\mysub{gbm,0}+226\unit{s}\), where \(T\mysub{gbm,0}\) is the trigger time reported by GBM. For the first \(4\unit{s}\) after the reference time (i.e., within \(230\unit{s}\) after the trigger), the TeV emission (if any) was only marginally detected with \lhaaso, implying a very rapid flux growth:
\be
F\mysub{0}\approx2\times10^{-6} \qty(\frac{t-T_*}{4\unit{s}})^{14.9^{+5.7}_{-3.9}}\ergscm\,.
\ee
We dub this part of the light curve as ``interval 0''. 
This rapid initial growth was followed with a slower rising phase \citep[interval ``\emph{a}'' from][]{2023Sci...380.1390.}, which lasted for \(\approx 14\unit{s}\):
\be
F\mysub{a}\approx2\times10^{-6} \qty(\frac{t-T_*}{4\unit{s}})^{1.8^{+0.21}_{-0.18}}\ergscm\,.
\ee
After that the flux saturated at the level of \(F\mysub{max}=F\mysub{b}\approx2\times10^{-5}\ergscm\) \citep[interval ``\emph{b}'' from][]{2023Sci...380.1390.}.
Following the peak, the light curve started its initial fading phase \citep[intervals ``\emph{c+d}'' from][]{2023Sci...380.1390.}. During \(\approx 10^3\unit{s}\), the TeV flux evolved as
\be
F\mysub{c}\approx2\times10^{-5} \qty(\frac{t-T_*}{18\unit{s}})^{-1.115^{+0.012}_{-0.012}}\ergscm\,.
\ee
Finally, after \(T_*+670_{-110}^{+230}\unit{s}\) this decay accelerated:
\be
F\mysub{e}\propto\qty({t-T_*})^{-2.21^{+0.30}_{-0.83}}\,,
\ee
with the index change of \(1.1^{+0.8}_{-0.3}\). We refer to this faster decay period as interval ``\emph{e}''.
\section{Discussion}
The detection of \grbLI with \lhaaso revealed a surprisingly smooth light curve. This light curve reveals several features that
constrain the parameters of this burst and can shed light on the afterglow physics of GRBs. These features include: (i)
the offset of the reference time with respect to the trigger time, (ii) a range of power-law indexes of the light curve,
(iii) positions of the breaks in these power-laws. 
We interpret the VHE light curve assuming that the jet interaction occurs either with the stellar wind or with homogeneous circumburst medium.

 We also note that the reported trigger time doesn't necessary determine the moment of the jet activation, which may be delayed with
respect to the trigger time. Thus, we assume that the physical trigger time, \(T_0\), is to some extent uncertain,
\(T_0>T\mysub{gbm,0}\). 

\subsection{Interaction with homogeneous circumburst medium}
\label{jet_homogeneous}

\begin{figure}
  \plotone{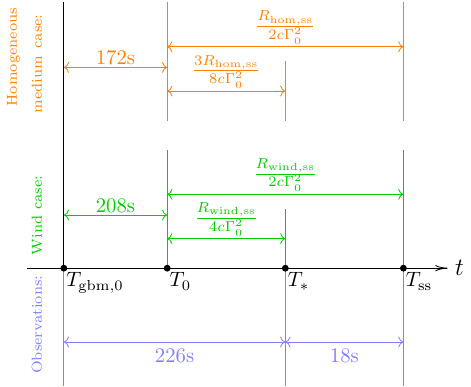}
  \caption{Relation between the key time-scale for a GRB jet interacting with progenitor wind and homogeneous medium shown together with the time-scales revealed with \lhaaso from \grbLI.\label{fig:time_scale}}
\end{figure}
The GRB jet interacting with homogeneous ISM was proposed to be the most natural explanation for the \lhaaso observations \citep{2023Sci...380.1390.}. The key feature that reportedly supports this scenario is the growth of the light curve during interval ``\emph{a}'', which was interpreted as emission from the coasting phase during which the blastwave propagates within a homogeneous ISM \citep{2023Sci...380.1390.}. In this case,  the transition to the self-similar regime occurs at the observer time \(T\mysub{ss}\approx R\mysub{h,ss}/(2c\Gamma_0^2)\) after the trigger time, and the offset of the reference time of the self-similar solution is \(T_*=3T\mysub{ss}/4\). Therefore the transition to the self-similar phase occurs at \(T\mysub{ss}\approx R\mysub{h,ss}/(8c\Gamma_0^2)\) after the reference time, and the jet activation (see in Fig.~\ref{fig:time_scale}) was at
\be
T_0\approx T_* - \frac34 T\mysub{ss}\approx T\mysub{gbm,0}+172\unit{s}\,.
\ee
Remarkably, at this moment, the main burst episode starts in the Fermi/GBM light curve. We therefore adopt this point as the jet activation moment, and one should set \(T\mysub{ss}\approx72\unit{s}\). The blast wave at this moment has a radius of 
\be\label{eq:Rss_ism}
R\mysub{ss}\approx 2cT\mysub{ss}\Gamma_0^2\approx4\times10^{17}\qty(\frac{\Gamma_0}{300})^2\qty(\frac{T\mysub{ss}}{72\unit{s}})\unit{cm}\,.
\ee
On the other hand, the transition implies that the ISM mass is
\be
M\mysub{ss}=\frac{4\pi}{3}R\mysub{ss}^3m_pn\approx \frac{E_0}{\Gamma_0^2 c^2}\,.
\ee
Solving these two equations we obtain an almost parameter-independent estimate for the initial Lorentz factor
\be\label{eq:gamma_ism}
\Gamma_0\approx600\qty(\frac{E_0}{10^{55}\unit{erg}})^{\nicefrac{1}{8}}\qty(\frac{n}{10^{-3}\unit{cm^{-3}}})^{\nicefrac{-1}{8}}\qty(\frac{T\mysub{ss}}{72\unit{s}})^{\nicefrac{-3}{8}}\,,
\ee
where we use a density value normalized to a value typical for the shocked stellar wind, Eq.~(\ref{eq:bbl_r_density}).

\begin{figure}
  \plotone{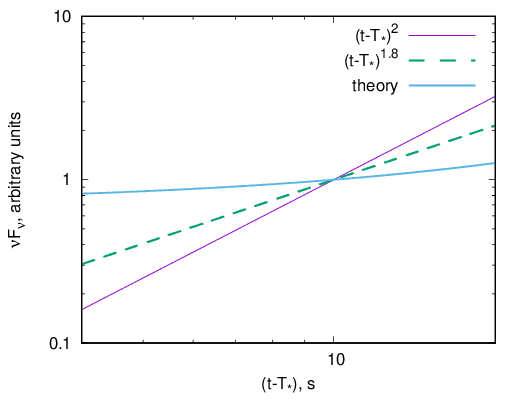}
  \caption{Change of the GRB luminosity caused by the shock dynamics  expected for the case of GRB jet interacting with homogeneous medium during the coasting phase, Eq.~\eqref{eq:coasting_ism_iso}, shown as a function of \(t-T_*\) and label ``theory''. The dependence is compared to \lhaaso interval ``\emph{a}'' (temporal index \(1.8\)) and to curve that ignores the influence of the offset (temporal index \(2\)). \label{fig:coasting_lum}}
\end{figure}

Which dependency of the shock luminosity should one expect on \((t-T_*)\) during the coasting phase? Theory predicts that the propagation of a jet with a constant Lorentz factor results in  a ``\(\propto t^2\)'' increase of the flux. This dependency, however, is for the detection time with respect to the trigger time. Thus, during the coasting phase theory predicts a flux dependence as
\be\label{eq:coasting_lum}
\begin{split}
F\mysub{coasting} &\propto \qty(t-T_*+54\unit{s})^2\\
&\propto\qty[\qty(\frac{t-T_*}{1\unit{s}})^2 +108\frac{\qty(t-T_*)}{1\unit{s}} + 3\times10^3]\,.
\end{split}
\ee
\lhaaso reported an index of \(1.8\) relative to the reference time for the time interval between \(4\) and \(18\unit{s}\) after the reference time. In Fig.~\ref{fig:coasting_lum} we plot in a log scale the dependency given by Eq.~\eqref{eq:coasting_lum}, and compare it to  \(\qty(t-T_*)^{1.8}\) and \(\qty(t-T_*)^2\). It can be seen from this figure that one expects an almost constant flux level during the coasting phase if plotted with respect to \(\qty(t-T_*)\), which significantly disagrees with the \lhaaso data. 

Thus, if the GRB jet indeed interacts with a homogeneous circumburst medium, then both the rapid increase of the TeV flux within the first \(4\unit{s}\) after $T_{*}$ (i.e., interval 0), and the slower flux increase during the next \(\sim15\unit{s}\) (i.e., interval ``\emph{a}'') must be caused by processes internal to the jet. For example, the following processes may result in such rapid flux growth: gamma-gamma absorption;  activation of the acceleration process; or development of the target photon field. Since the attenuation as well as acceleration activation have a very strong impact on the flux level, the relatively smooth flux increase between \(T_*+4\unit{s}\) and \(T_*+18\unit{s}\) appears to be more naturally explained by the development of the target, and the more abrupt initial increase, \(t<T_*+4\unit{s}\), by the decrease of gamma-gamma absorption or by activation of the acceleration mechanism.

Assuming that the development of the photon target explains the growth of the VHE emission before \(t-T_*\approx18\unit{s}\) (during interval ``a"), using Eq.~\eqref{eq:tau_target} for \(t\mysub{ph}\approx70\unit{s}\), one can obtain an estimate for the required magnetic field
\be
B'\approx 0.2\qty(\frac{\hbar\omega}{1\unit{TeV}})^{\nicefrac{1}{3}}\qty(\frac{E_0}{10^{55}\unit{erg}})^{\nicefrac{-1}{8}}\times
\ee
\[
  \qty(\frac{n}{10^{-3}\unit{cm^{-3}}})^{\nicefrac{1}{8}}\qty(\frac{T\mysub{ss}}{72\unit{s}})^{\nicefrac{3}{8}}\unit{G}\,.
\]
This result only weakly depends on all uncertain parameters. For the typical parameter values usually adopted, this magnetic field strength corresponds to a magnetization of \(\sim3\times10^{-3}\). We also note that development of IC target should be accompanied by hardening of the IC spectrum \citep[see][and discussion in Sec.~\ref{sec:target}]{2023arXiv230712467K}, which appears consistent with \lhaaso observations \citep[see in Fig. 3 of][]{2023Sci...380.1390.}.

The gamma-gamma optical depth depends on the emitter size, its bulk Lorentz factor, and density of the target photons. Since we have constrained both the Lorentz factor and the shock radius, we can estimate the luminosity of the target photon field required to lead to considerable attenuation during the interval ``0'' and compare it to the available observational data. According to the results obtained with Konus-Wind instrument \citep{2023ApJ...949L...7F}, the energy flux carried by \(20\unit{keV}\)~--~\(10\unit{MeV}\) photons was \(\approx\qty(1.62\pm0.09)\times10^{-2}~\ergs\unit{cm^{-2}}\)  between \(T\mysub{gbm,0}+225\unit{s}\) to \(T\mysub{gbm,0}+233\unit{s}\). This energy flux corresponds to \(\approx10^{54}~\ergs\) for \(z=0.151\). The part suitable for attenuation of TeV photons is likely \(\sim10^{52}~\ergs\). Since this value seems to be (marginally) comparable to the estimate given by Eq.~\eqref{eq:atten_lum}, one cannot rule out that the change of gamma-gamma absorption is reflected in the light curve. 

Using the Lorentz factor given by Eq.~\eqref{eq:gamma_ism}, Eq.~\eqref{eq:Rss_ism} implies the radius for which the coasting phase ends is at \(\lesssim \unit{pc}\). This distance is likely smaller than the inner size of the stellar bubble, thus the jet~--~homogeneous medium regime may be realized only in the case of a very weak wind of the GRB progenitor. More specifically this scenario requires a stellar bubble of outer radius of \(\approx 10\unit{pc}\), which requires a stellar wind kinetic luminosity of \(L\mysub{w}\sim 10^{33}\ergs\), see Eq.~\eqref{eq:stellar_bubble}.

If the GRB jet interacts with a homogeneous medium, then, using the estimate given by Eq.~\eqref{eq:radius_bbl}, the GRB forward shock should reach a distance of a few parsec  at \(t-T_*\approx670\unit{s}\). We do not expect any considerable change of the shock dynamics at this point, since the compressed ISM layer should be located at a few tens of parsec (unless the stellar wind kinetic luminosity is tiny, \(L\mysub{w}< 10^{32}\ergs\)).

Thus, the change of the light curve power-law index at \(T_*+670\unit{s}\) (i.e., the transition to interval ``\emph{e}'') is most likely caused by the jet breaking \citep[see][for the implications of this scenario, however, note the difference in homogeneous medium density]{2023Sci...380.1390.}. The change of temporal index of \(1.1^{+0.8}_{-0.3}\) is marginally consistent with the change of \(0.75\) predicted by theory (i.e., the temporal index is expected to change from \(-1.12\) to \(-1.87\approx-1.9\)). Thus, further detailed simulations are needed to test the feasibility of this scenario.

\subsection{Interaction with wind}
\label{jet_wind}

As shown in section~\ref{jet_homogeneous}, the \lhaaso light curve does not favor the scenario of interaction with homogeneous circumburst medium. Instead, the interaction with progenitor wind is found to provide an equally feasible scenario. In this case, the transition to the self-similar phase and the offset of the reference time differ by a factor of \(2\). Thus, the jet activation time should occur at  (see in Fig.~\ref{fig:time_scale})
\be
T_0\approx T_* - \frac{R\mysub{w,ss}}{4c\Gamma_0^2} \approx T\mysub{gbm,0}+208\unit{s}\,.
\ee
In the GBM light curve this moment approximately corresponds to the onset of the main burst, that saturated the instrument. Thus, this can be considered an indirect support for this scenario.

The radius of the blast wave at this moment is  
\be\label{eq:Rss_wind}
R\mysub{ss}\approx 2cT\mysub{ss}\Gamma_0^2\approx2\times10^{17}\qty(\frac{\Gamma_0}{300})^2\qty(\frac{T\mysub{ss}}{36\unit{s}})\unit{cm}\,.
\ee
The transition to the self-similar regime implies that the mass of the shocked wind is 
\be
M\mysub{ss}=\frac{\dot{M}\mysub{w} R\mysub{ss}}{v\mysub{w}}\approx \frac{E_0}{\Gamma_0^2 c^2}\,.
\ee
Solving these two equations for the initial Lorentz factor we obtain
\be\label{eq:G0_wind}
\Gamma_0\approx600\qty(\frac{E_0}{10^{55}\unit{erg}})^{\nicefrac{1}{4}}\qty(\frac{v\mysub{w}}{2\times10^3\unit{km\,s^{-1}}})^{\nicefrac{1}{4}}\times
\ee
\[
  \qty(\frac{\dot{M}\mysub{w}}{10^{-7}\mathrm{M}_\odot\unit{yr^{-1}}})^{\nicefrac{-1}{4}}\qty(\frac{T\mysub{ss}}{36\unit{s}})^{\nicefrac{-1}{4}}\,.
\]
Substituting Eq.~\eqref{eq:G0_wind} to Eq.~\eqref{eq:Rss_wind}, one obtains the distance at which the interaction enters the self-similar regime. For the adopted parameter values, it should be a sub-parsec distance, i.e., well inside the hot stellar bubble.

If the emission is generated at the jet interaction with the stellar wind, then the \((t-T_*)^{1.8}\) part of the light curve (i.e., interval ``\emph{a}'') should be caused by processes internal to the jet. Similar arguments as in the previous case considered also apply here, thus the processes related to the development of the target are the most feasible explanation for this phase. In particular, Eq.~\eqref{eq:tau_target} can explain this growth phase if 
\be
B'\approx 0.3\qty(\frac{\hbar\omega}{1\unit{TeV}})^{\nicefrac{1}{3}}\qty(\frac{E_0}{10^{55}\unit{erg}})^{\nicefrac{-1}{4}}\times
\ee
\[
  \qty(\frac{v\mysub{w}}{2\times10^3\unit{km\,s^{-1}}})^{\nicefrac{-1}{4}}\qty(\frac{\dot{M}\mysub{w}}{10^{-7}\mathrm{M}_\odot\unit{yr^{-1}}})^{\nicefrac{1}{4}}\qty(\frac{T\mysub{ss}}{36\unit{s}})^{\nicefrac{1}{4}}\unit{G}\,.
\]
For the typical parameter values, this magnetic field strength corresponds to a magnetization of \(\sim3\times10^{-2}\).

If \grbLI is produced by a jet interacting with the stellar wind, then the initial rapid increase of the TeV flux (i.e., interval 0) can be attributed to the activation of the acceleration mechanism, e.g., to the magnetic field amplification, or to the impact of gamma-gamma absorption. We note that because of the different relation between the trigger and reference times, in the progenitor wind scenario the shock locates closer to the explosion origin and thus the impact of gamma-gamma absorption is stronger compared to the homogeneous circumburst medium case. However, this impact needs to be verified with more accurate calculations (to be presented elsewhere). 

\begin{figure}
  \plotone{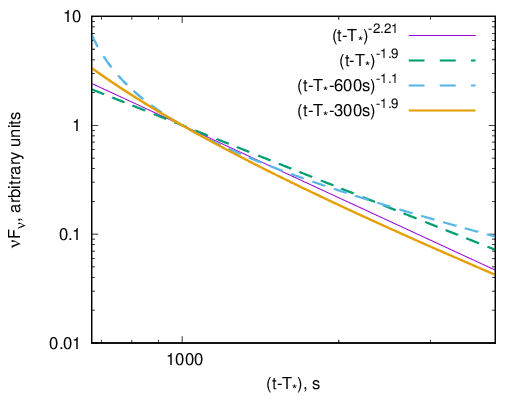}
  \caption{GRB luminosity measured with \lhaaso during the interval ``\emph{e}'' compared to theory predicted temporal evolution after jet break (temporal index \(-1.12-0.75\approx-1.9\)), expected during the shock propagating in the bubble (additional offset); and jet break occurring in the bubble (temporal index \(-1.9\) and additional offset).\label{fig:bubble_lum}}
\end{figure}

If the GRB jet interacts with the progenitor wind, then the break at \(T_*+670\unit{s}\) (i.e., the transition to interval ``\emph{e}'') can be caused by the jet break. The index change of \(1.1^{+0.8}_{-0.3}\) seems to be inconsistent with the change of \(0.5\) predicted by the theory for the jet  breaking in the stellar wind. On the other hand, according to Eq.~\eqref{eq:radius_wind} the forward shock is expected to reach a distance of \(R\mysub{br}\approx2\unit{pc}\) at \(T_*+670\unit{s}\). This length matches quite well the inner size of the stellar bubble. Thus it seems feasible that the transition to the interval ``\emph{e}'' is instead related to the blast wave interaction with the inner boundary of the stellar bubble. There are a few effects that need to be taken into account to verify this hypothesis. In the first place, once the jet reaches the stellar bubble it starts interacting with homogeneous medium, so its dynamics changes (see Appendix \ref{app:bubble} for detail). Another point is related to the change of the reference time expected after the blast wave adjusts to the new propagation regime. As discussed in Sec.~\ref{sec:bbl}, the reference times for the self-similar expansion of the blast wave in the wind zone and in the bubble differ. Similarly to the case summarized in Fig.~\ref{fig:coasting_lum}, this change of the reference time may cause a significant distortion of the light curve when it is plotted in a log scale with respect to a different reference time. In Fig.~\ref{fig:bubble_lum} we present some examples that illustrate this effect. In interval ``\emph{e}'' \lhaaso obtained a time-dependence \((t-T_*)^{-2.21}\), which is faster than the dependence predicted by theory for the jet break scenario, \((t-T_*)^{-1.9}\). If, however, one assumes an additional offset of the reference time (which is justified by the analysis presented in Sec. \ref{sec:bbl}), then the theoretical predictions match the observations better, see the case \((t-T_*-300\unit{s})^{-1.9}\) in Fig.~\ref{fig:bubble_lum}. We note that this case corresponds to the jet break occurring in the shocked stellar wind. Finally, if the additional offset of the reference time is significant, then it causes a sharp transformation of the light curve (see the case \((t-T_*-600\unit{s})^{-1.1}\)), which could alleviate the need for a jet break. However, a detailed light curve modeling is required to achieve any robust conclusion here. 

\section{Summary}

The detection of \grbLI with \lhaaso provides a unique data sample that allows a very insightful analysis of the processes occurring in the early afterglow phase. In this paper we have presented a qualitative study of the key features of the light curve obtained with \lhaaso and have shown that this data set allows one to constrain the key parameters of the burst, in particular, the jet activation moment, its initial Lorentz factor, and magnetic field strength. The obtained parameter values of the initial Lorentz factor and magnetic field strength agree with the ones typically revealed by spectral modeling of GRBs, and the jet activation moment obtained solely based on properties of the VHE light curve matches a very special point in the GBM light curve. This remarkable match can be considered as an indirect confirmation of the considered scenario, and as a support for the GRB phenomenology, in general.

We find that whilst the \lhaaso data do not exclude the homogeneous circumburst medium scenario, the progenitor wind scenario looks preferable as it appears to show excellent agreement with the expected size and structure of the stellar bubble.

The apparent agreement of the power-law slopes characterizing the light curve obtained with \lhaaso (i.e., a change of its power-law index from \(+1.8\) to \(-1.115\) to \(-2.21\)) with the predictions for a GRB jet interacting with homogeneous circumburst medium was considered as a strong support for this scenario \citep{2023Sci...380.1390.}. The analysis performed here, however, suggests that one needs to reconsider the validity of such an argument. Indeed, for an explosion into a homogeneous medium we expect a growth of the flux, \(\propto t^2\), during the coasting phase. However, this dependence appears as an almost constant flux, if plotted on a log scale, with respect to the reference time of the self-similar phase.
%
Therefore, we cannot give preference to either one of the two standard scenarios, jet interaction with a homogeneous circumbust medium or jet interaction with the progenitor wind, solely based on interval ``\emph{a}'' of the \lhaaso light curve.
We also note that provided the early nature of the afterglow detected with \lhaaso, the hot stellar bubble should be considered as the homogeneous medium interacted with, not the standard ISM.

{\bf Homogeneous medium scenario.} If the homogeneous medium  scenario is realized, then the light curve implies an initial Lorentz factor, \(\Gamma_0\approx 600\). Also, the emission should be generated at \(<\unit{pc}\) from the explosion origin, i.e., from a region expected, for the standard parameter values, to be well within the wind zone. Thus, the homogeneous medium scenario requires a weak progenitor wind.

The initial rapid increase of the VHE flux (i.e., interval 0) can be explained by either a weakening of the gamma-gamma attenuation on the photons from the prompt emission, or by delay due to the activation time of the acceleration process. Smoother increase between \(4\unit{s}<(t-T_*)<18\unit{s}\) (i.e., interval ``\emph{a}'') can be explained by development of the photon target for SSC process, if the magnetic field in the production regions is \(B'\approx 0.2\unit{G}\), which corresponds to quite a small magnetization in the production region, \(\sim3\times10^{-3}\).

Finally, the softening of the light curve at \(T_*+670\unit{s}\) (i.e., the transition to interval ``\emph{e}'') can be explained by a jet break as suggested in the discovery paper \cite{2023Sci...380.1390.} 

or by the blast wave reaching the contact discontinuity, where the medium density undergoes an almost a \(10^3\) fold increase. The latter explanation, however, is found to require an unrealistic constraint on the wind kinetic luminosity. 

{\bf Wind scenario.} On the other hand, the wind scenario implies a similar initial Lorentz factor, \(\Gamma_0\approx600\). The location of the interaction region, \(\sim10^{17}\unit{cm}\), is smaller than the inner size of the stellar bubble for typical values of the parameters, thus the wind scenario looks preferable from this perspective.

Similar to the homogeneous circumburst medium case, the initial rapid growth (i.e., interval 0) can be caused by gamma-gamma absorption or by the activation of the acceleration process. The smoother increase seen in the light curve during interval ``\emph{a}'' can be explained by the development of the SSC target, if the magnetic field is \(B'\approx0.3\unit{G}\), which translates to a \(\sim3\times10^{-2}\) magnetization of the forward shock downstream (in the homogeneous circumburst medium case, a similar magnetic field strength points to a lower value of the magnetization because of the expected higher density of upstream medium). The break in light curve seen at \(T_*+670\unit{s}\) (i.e., the transition to interval ``\emph{e}'') appears to be too strong to be consistent with a jet break expected in the wind scenario (the index change of \(1.1^{+0.8}_{-0.3}\) instead of theory predicted change of \(0.5\)). This could be taken as an indication for a change of the jet dynamics caused by the interaction with the stellar bubble at \(R\approx 2\unit{pc}\). We note that this transition also causes an addition offset of the reference time, which alone can lead to an apparent break in the light curve. 

\begin{figure}
  \plotone{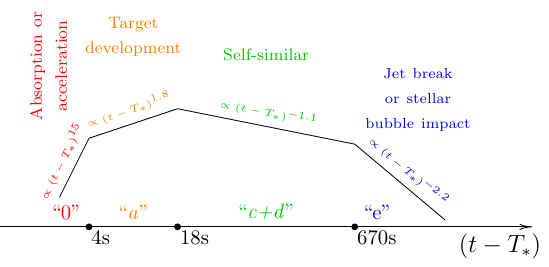}
  \caption{Different phases of the early afterglow shown with labels indicating the key physical processes responsible for the light curve evolution.\label{fig:lightcuve}}
\end{figure}

A summary of the physical processes responsible for the formation of the VHE light curve are shown in Fig.~\ref{fig:lightcuve}. In both scenarios we conclude that the growing part of the light curve is dominated by processes internal to the jet (including the gamma-gamma attenuation on photons from the prompt phase) and that the decaying parts are due to the jet dynamics, namely the jet propagation in the self similar regime and jet breaking. Although the homogeneous medium case cannot be excluded based on qualitative analysis presented here, the progenitor wind scenario is favoured due to the inferred length scales naturally fitting the expected size of the wind zone and the stellar bubble.

The Lorentz factor and the magnetic field derived merely from the analysis of the light curve,  significantly reduce the parameter space for modeling the time-dependent SED of the afterglow. The results of the modeling of the synchrotron-self-Compton SED taking into account the internal absorption, will be published elsewhere.

\begin{acknowledgments}
D.K. acknowledges the support of RSF grant No. 21-12-00416. A.T. acknowledges support from DESY (Zeuthen, Germany), a member of the Helmholtz Association HGF.
\end{acknowledgments}
\appendix

\section{Shock wave transition from the wind zone to the stellar bubble}\label{app:bubble}
When the shock wave reaches the inner boundary of the stellar bubble, one expects a change of the propagation regime as the upstream media density changes from \(1/R^2\) to \(1/R^0\) (i.e., constant) dependency. The density of the upstream homogeneous medium is only a factor of \(4\) larger than the stellar wind density at the wind termination shock. It means that the dynamics of the shock close to the termination shock is influenced by mass accumulated in the wind zone considerably, thus one needs to account for this contribution.

Density of the upsteam gas is then
\be
n=\left\{
\begin{matrix}
    \frac{\dot{M}\mysub{w}}{4\pi R^2 m_pv\mysub{w}}&\qq{for}&R<R\mysub{bbl,r}\,,\\
    n\mysub{bbl,r}&\qq{for}&R> R\mysub{bbl,r}\,,
\end{matrix}
\right.
\ee
where from the shock jupm condition we have
\be
n\mysub{bbl,r}=\frac{\dot{M}\mysub{w}}{\pi R\mysub{bbl,r}^2 m_pv\mysub{w}}\,.
\ee
Therefore the mass of the shocked shell depends on \(R\) as
\be
M=\left\{
\begin{matrix}
    \frac{R\dot{M}\mysub{w}}{v\mysub{w}}&\qq{for}&R<R\mysub{bbl,r}\\
    \frac{R\mysub{bbl,r}\dot{M}\mysub{w}}{v\mysub{w}}+\frac{4\pi}{3}n\mysub{bbl,r}m_p\qty(R^3-R\mysub{bbl,r}^3)&\qq{for}&R\geq R\mysub{bbl,r}
\end{matrix}
\right.
\ee
Using the standard relation between the shell mass and bulk Lorentz factor, one obtains
\be
\Gamma^2=\frac{E_0}{c^2}\times\left\{
\begin{matrix}
    \frac{v\mysub{w}}{R\dot{M}\mysub{w}}&\qq{for}&R<R\mysub{bbl,r}\\
    \qty[\frac{R\mysub{bbl,r}\dot{M}\mysub{w}}{v\mysub{w}}+\frac{4\pi}{3}n\mysub{bbl,r}m_p\qty(R^3-R\mysub{bbl,r}^3)]^{-1}&\qq{for}&R\geq R\mysub{bbl,r}
\end{matrix}
\right.
\ee

The relation between the blast wave radius and the delay can be obtained with simple integrations:
\be\label{eq:apendix_t_on_R}
t=\left\{
\begin{matrix}
    \frac{R}{2c\Gamma_0^2}&\qq{for}& R\leq R\mysub{w,ss}\\
    \frac{R\mysub{w,ss}}{2c\Gamma_0^2}+\frac{\dot{M}\mysub{w}\qty(R^2-R\mysub{w,ss}^2)}{4cE_0v\mysub{w}}&\qq{for}& R\mysub{w,ss}< R\leq R\mysub{bbl,r}\\
    t\qty(R\mysub{bbl,r})+\frac{\dot{M}\mysub{w}cR\mysub{bbl,r}\qty(R-R\mysub{bbl,r})}{2E_0v\mysub{w}}+ \frac{\pi}{6}\frac{n\mysub{bbl,r}m_pc\qty(R^4-R\mysub{bbl,r}^4)}{E_0}- \frac{4\pi}{6}\frac{n\mysub{bbl,r}m_pc\qty(R-R\mysub{bbl,r})R\mysub{bbl,r}^3}{E_0}&\qq{for}&R>R\mysub{bbl,r}
\end{matrix}
\right.
\ee
However, for the purpose of interpreting observations the inverse dependence, i.e. \(R(t)\), is required. The first two cases from the above equations are trivial: 
\be
R=\left\{
\begin{matrix}
    2c\Gamma_0^2 t&\qq{for}& t\leq \frac{R\mysub{w,ss}}{2c\Gamma_0^2}\,,\\
    \sqrt{\frac{4cE_0v\mysub{w}\qty(t-T_*)}{\dot{M}\mysub{w}}}&\qq{for}& \frac{R\mysub{w,ss}}{2c\Gamma_0^2}<t\leq t\qty(R\mysub{bbl,r})\,,\\
    \tilde{R}(t)&\qq{for}&t>t\qty(R\mysub{bbl,r})\,,
\end{matrix}
\right.
\ee
and the last one, \(t>t\qty(R\mysub{bbl,r})\), requires some simple algebra. Substituting the bubble density in to the third equation of Eq.~\eqref{eq:apendix_t_on_R} one obtains:
\be\label{eq:appendix_xy}
y^4-y = x\,,
\ee
where
\be
\begin{split}
  y&=\frac{R}{R\mysub{bbl,r}}\,,\\
  x&=\frac{6E_0v\mysub{w}}{\dot{M}\mysub{w}cR\mysub{bbl,r}}\qty(t-t\qty(R\mysub{bbl,r}))\,.
\end{split}
\ee
Equation~\eqref{eq:appendix_xy} is polynomial equation of \(4^{\mathrm{th}}\) power, thus it allows analytical solution. The physical root of the equation can be selected by the condition \(\eval{y}_{x=0}^{}=1\):
\be
\begin{split}
y =& {{\sqrt{3\,\left({{
 \sqrt{256\,x^3+27}}\over{2\,3^{{{3}\over{2}}}}}+{{1}\over{2}}\right)
 ^{{{2}\over{3}}}-4\,x}}\over{2\,\sqrt{3}\,\left({{\sqrt{256\,x^3+27}
     }\over{2\,3^{{{3}\over{2}}}}}+{{1}\over{2}}\right)^{{{1}\over{6}}}}} + \\
  & {{\sqrt{{{2\,\sqrt{3}\,\left({{\sqrt{256\,x^3+27}}\over{2\,3^{{{3
 }\over{2}}}}}+{{1}\over{2}}\right)^{{{1}\over{6}}}}\over{\sqrt{3\,
 \left({{\sqrt{256\,x^3+27}}\over{2\,3^{{{3}\over{2}}}}}+{{1}\over{2
 }}\right)^{{{2}\over{3}}}-4\,x}}}-\left({{\sqrt{256\,x^3+27}}\over{2
 \,3^{{{3}\over{2}}}}}+{{1}\over{2}}\right)^{{{1}\over{3}}}+{{4\,x
 }\over{3\,\left({{\sqrt{256\,x^3+27}}\over{2\,3^{{{3}\over{2}}}}}+{{
 1}\over{2}}\right)^{{{1}\over{3}}}}}}}\over{2}}\,.
\end{split}
\ee
%
The asymptotic behavior of this expression is
\be
y = \left\{
  \begin{matrix}
    1 + \frac{x}{3} -\frac{2x^2}{9}+\frac{20x^3}{81}&\qq{for}& x\ll 1\\
    x^{\nicefrac{1}{4}}+\frac{1}{4}x^{\nicefrac{-1}{4}}&\qq{for}&x\gg 1
  \end{matrix}\,.
\right.
\ee
%




\end{document}